\documentclass[conference]{IEEEtran}
\IEEEoverridecommandlockouts
\usepackage{cite}
\usepackage{amsmath,amssymb,amsfonts}
\usepackage{algorithmic}
\usepackage{graphicx}
\usepackage{textcomp}
\usepackage{xcolor}
\usepackage{hyperref}       % hyperlinks
\usepackage[caption=false, font=footnotesize]{subfig}
\def\BibTeX{{\rm B\kern-.05em{\sc i\kern-.025em b}\kern-.08em
    T\kern-.1667em\lower.7ex\hbox{E}\kern-.125emX}}

\usepackage{url}
\usepackage{times}
\usepackage{ifthen}   % for conditionals
\usepackage{algorithm}
\usepackage{etoolbox}
\usepackage{algorithmic}
\usepackage{color}
\usepackage{multirow}
\usepackage{mathabx}
\usepackage{longtable} 	
\usepackage{datetime}
\usepackage{xspace}
\usepackage{paralist} %% This gives enumeration environments & compactenum 
\usepackage{enumitem} % For tighter lists
\usepackage{cleveref} % for cref

\usepackage{titlesec}
\usepackage{mdframed}  % creating color text boxes
\usepackage{colortbl}
\usepackage{soul}

\makeatletter
\def\namedlabel#1#2{\begingroup
    #2%
    \def\@currentlabel{#2}%
    \phantomsection\label{#1}\endgroup
}
\makeatother

\usepackage{bm} % For bold math symbols

\newcommand{\thetab}{\bm{\theta}}
\newcommand{\etab}{\bm{\eta}}
\newcommand{\epsb}{\bm{\varepsilon}}

\newcommand{\yb}{\bm{y}}

\newcommand{\etabo}{\bm{\eta}^{\text{obs}}}
\newcommand{\etaboT}[2]{\bm{\eta}^{\text{obs}}_{#1:#2}}
\newcommand{\etabT}[2]{\bm{\eta}_{#1:#2}}
\newcommand{\ybT}[2]{\bm{y}_{#1:#2}}

\usepackage{tikz}
\usetikzlibrary{positioning}
\usetikzlibrary{calc}
\usetikzlibrary{shapes.geometric}
\usetikzlibrary{topaths}
\usetikzlibrary{external}
\usetikzlibrary{fadings,decorations.pathreplacing} 

\usepackage{pgfplots}
\usetikzlibrary{shapes,decorations.pathmorphing,patterns}
\pgfplotsset{compat=newest}
\usepgfplotslibrary{fillbetween}

\usepackage[flushleft]{threeparttable}
\begin{document}
% Towards sequential monte carlo uncertainty \\ time-varying 
\title{Towards Improved Uncertainty Quantification of Stochastic Epidemic Models Using\\ Sequential Monte Carlo
\thanks{This material is based upon work supported by the National Science
Foundation under Grants 2200234 and CCF-1918770, the U.S. Department of Energy, Office of Science, under contract number DE-AC02-06CH11357 and the Bio-preparedness Research Virtual Environment (BRaVE) initiative. This research was completed with resources provided by the Laboratory Computing Resource Center at Argonne National Laboratory.}
}

% put name of the funding agencies here to be acknowledged
% NSF PIPP-1
% DOE ASCR BRaVE1
% NSF Expeditions Grant CCF-1918770 (Dave)

\author{
\IEEEauthorblockN{
Arindam Fadikar\IEEEauthorrefmark{1},
Abby Stevens\IEEEauthorrefmark{1},
Nicholson Collier\IEEEauthorrefmark{1},
Kok Ben Toh\IEEEauthorrefmark{2} \\
Olga Morozova\IEEEauthorrefmark{3},
Anna Hotton\IEEEauthorrefmark{4},
Jared Clark\IEEEauthorrefmark{5},
David Higdon\IEEEauthorrefmark{5} and
Jonathan Ozik\IEEEauthorrefmark{1}} \IEEEauthorblockA{\IEEEauthorrefmark{1}Decision and Infrastructure Sciences\\
Argonne National Laboratory, Lemont, IL, USA \\
Email: afadikar@anl.gov, stevensa@anl.gov, ncollier@anl.gov, jozik@anl.gov} 
\IEEEauthorblockA{\IEEEauthorrefmark{2}Department of Preventive Medicine \\
Northwestern University, Chicago, IL, USA \\
Email: bentoh@northwestern.edu}
\IEEEauthorblockA{\IEEEauthorrefmark{3}Department of Public Health Sciences \\
University of Chicago, Chicago, IL, USA \\
Email: omorozova@bsd.uchicago.edu}
\IEEEauthorblockA{\IEEEauthorrefmark{4}Department of Medicine \\
University of Chicago, Chicago, IL, USA \\
Email: ahotton@bsd.uchicago.edu}
\IEEEauthorblockA{\IEEEauthorrefmark{5}Department of Statistics \\
Virginia Tech, Blacksburg, VA, USA \\
Email: cjared96@vt.edu, dhigdon@vt.edu}

}

\maketitle

\pagestyle{plain}

\begin{abstract}
    Sequential Monte Carlo (SMC) algorithms represent a suite of robust computational methodologies utilized for state estimation and parameter inference within dynamical systems, particularly in real-time or online environments where data arrives sequentially over time. In this research endeavor, we propose an integrated framework that combines a stochastic epidemic simulator with a sequential importance sampling (SIS) scheme to dynamically infer model parameters, which evolve due to social as well as biological processes throughout the progression of an epidemic outbreak and are also influenced by evolving data measurement bias. Through iterative updates of a set of weighted simulated trajectories based on observed data, this framework enables the estimation of posterior distributions for these parameters, thereby capturing their temporal variability and associated uncertainties. Through simulation studies, we showcase the efficacy of SMC in accurately tracking the evolving dynamics of epidemics while appropriately accounting for uncertainties. Moreover, we delve into practical considerations and challenges inherent in implementing SMC for parameter estimation within dynamic epidemiological settings, areas where the substantial computational capabilities of high-performance computing resources can be usefully brought to bear.
    \vspace{0.1in}
    
\end{abstract}

\begin{IEEEkeywords}
sequential Monte-Carlo, computational epidemiology, uncertainty quantification
\end{IEEEkeywords}

\section{Introduction}
\label{sec:intro}

Computational epidemiologic models have become essential tools for public health decision-making, enabling rapid responses during times of crisis and uncertainty \cite{Ozik2021}. While the COVID-19 pandemic highlighted the possibilities enabled by computational epidemiological modeling, such as testing the impact of different interventions \cite{hotton_impact_2022}, it also revealed key weaknesses in modeling workflows \cite{collier_developing_2023}. Models are imperfect representations of the true epidemiological system, and for them to be used as trusted \textit{in silico} laboratories they need to be calibrated to empirical data, often from heterogeneous sources with varying degrees of uncertainty, while accounting for time-varying dynamics of an ongoing epidemic. 

Models themselves are often computationally demanding, and calibration, which typically requires repeated runs of a model, is even more so and necessitates the use of high-performance computing (HPC) resources. Furthermore, epidemics are constantly changing and evolving -- for example,  disease transmission may increase or decrease with time due to social and biological processes, and observational data, such as case counts (which are often used as a proxy for underlying infection incidence), may be reported at a rate that changes as a epidemic evolves in response to interventions or behavioral changes \cite{miller_reliability_2022}. 

Stochastic disease simulators, such as compartmental or agent-based models, further complicate parameter estimation by introducing additional randomness. Unlike deterministic models, which generate a fixed response given a particular input, multiple runs of a stochastic model with the same input parameters will generate an ensemble of random realizations of output trajectories. Traditional calibration techniques \cite{Kennedy2001} do not generally account for such randomness, hence so-called \textit{trajectory-oriented} optimization approaches have been developed to include the random seed as a parameter to be estimated \cite{fadikar2023trajectory}. Calibrating to individual trajectories, rather than summary statistics of the ensemble of outputs, can capture specific population dynamics, such as mixing patterns, and enables detailed scenario modeling and forecasting. 

In light of the described complexities and in an effort to advance the application of epidemiologic modeling for supporting robust public health decision making, we propose an HPC-aware framework in which stochastic simulators are calibrated sequentially against multiple updating data streams. In addition to estimating epidemic model parameters, we introduce a binomial bias model for the empirical data to account for reporting errors and additionally estimate this parameter. Using Sequential Importance Sampling (SIS, \cite{doucet2001sequential, cappe2007overview, papaioannou_sequential_2016}), we calibrate simulations within a given time frame against one or more empirical data sources and select the best trajectories. As the epidemic evolves and additional data become available, we calibrate the next time frame using parameter estimates from the previous time period as priors. Our system can be decomposed into an inner and outer loop corresponding to empirical data update cadences (e.g., daily, weekly), with the outer loop moving the model forward in time and the inner loop calibrating individual time periods. Estimating parameters sequentially in time allows us to capture their time-varying dynamics, as calibrating each time period results in different parameter estimates. 

By considering individual trajectories (rather than aggregations of stochastic output), we are able to store, or checkpoint, the exact state of the model, allowing models to be restarted from time-stamped stored states rather than restarting them from the beginning of an epidemic. This facilitates significant computational savings when deploying calibration algorithms in sequentially updating operational modes. Even with checkpointing facilities in place, producing sufficient parameter space samples from stochastic simulators can be computationally burdensome and thus our framework is designed to exploit the concurrency provided by HPC resources to make the computation feasible.

We demonstrate our proposed sequential inference procedure on a stochastic SEIR compartmental model that simulates COVID-19 transmission in Chicago \cite{runge2022modeling}. The model parameter of interest is the transmission rate, which we vary across time intervals. We simulate empirical case counts and deaths by applying a time-varying reporting bias to individual trajectories of the model. Our goal is to infer both the time-varying transmission rate and reporting bias using our SIS framework, and we calibrate to both case counts alone and a combination of case counts and deaths. 
% SAY SOMETHING ABOUT RESULTS HERE

Our framework is particularly suited for rapid public health decision-making -- it is computationally efficient, accounts for the uncertainties from various data sources, and can give up-to-date insights into the evolution of the epidemic. The rest of this paper proceeds as follows. In section \ref{sec:related}, we give an overview of related work and highlight novel elements of our system. Section \ref{sec:covid_age} describes the specific stochastic epidemic simulator we use to demonstrate our calibration method and checkpointing procedure. Section \ref{sec:method} formally defines the statistical model and derives the updates for our SIS procedure, and section \ref{sec:results} provides results from our experiment. The discussion can be found in section \ref{sec:discussion}.

\section{Related Work}
\label{sec:related}

% De-biasing references from Olya
% Sequential stuff
% Jeffrey Shaman @ columbia has something on data assimilation with epidemiological models
% 

%  olya has a set of papers that does debiasing
% Papers that do not-trajectory oriented optimization

 The development of Bayesian calibration methods for stochastic computational models, particularly for epidemic applications has seen significant progress in the recent past \cite{Fadikar2018, andrianakis2015bayesian, henderson2009bayesian}. However sampling-based, or Monte Carlo-based approaches for sensitivity analysis \cite{sobol2001global} or inference \cite{cornuet2008inferring, flury2011bayesian} on stochastic models is too demanding when running the simulations is expensive. In such cases, emulation – modeling the input-output response surface – has proven to be an effective approach. In some applications, where only the mean (or other summary statistics) response is modeled as a function of the input parameters \cite{Fadikar2018, kleijnen2009kriging, andrianakis2015bayesian}, the realizations from a simulation can be averaged, preprocessed, or accounted for in the inference framework. In the case of an agent-based epidemic model where each realization may indicate a specific mixing behavior among the agents, it is not well understood what the average (or other summary) of such mixing behavior would represent \cite{fadikar2023trajectory}. This issue further hinders the ability to use a \textit{mean} calibrated epidemic model to simulate forward in time. 

% Papers Olya sent (this is clunky):
Understanding epidemic dynamics in the face of incomplete or biased data and imperfect computational models has long been of interest to the scientific community, and research in this space has proliferated since the start of the COVID-19 pandemic.
Many have studied the impact of incomplete or biased data on the estimation of infection rates (e.g., \cite{gamado2014modelling, pitzer2021impact}) and the effective reproductive number $R_t$ \cite{gostic2020practical, hettinger2023estimating, white2009estimation, white2010reporting}. While most existing studies consider mechanistic models, such as compartmental or agent-based models, attempts to bridge traditional epidemiology with more recent statistical methods have shown promise in addressing data gaps \cite{quick2021regression, shi2022robust, white2021statistical, parag2022quantifying}. 
Similar to our work, \cite{Spannaus2022} estimate time-varying epidemic parameters and case reporting errors using a Sequential Monte-Carlo calibration approach; however, they cite computation as a major bottleneck in their analysis, which we address with checkpointing and exploiting the concurrency in high-performance computing systems. Furthermore, our work seeks to extend on other efforts to operationalize epidemiological modeling systems, such as \cite{shaman_real-time_2013} who developed real-time influenza forecasts using advanced data assimilation techniques.

\section{Stochastic Disease Simulator}
\label{sec:covid_age}

We utilize a stochastic SEIR compartmental model published by \cite{runge2022modeling} that simulates the transmission of SARS-CoV-2 and the progression of COVID-19 disease states to illustrate our proposed sequential inference procedure. This model incorporates various symptom statuses, including asymptomatic, presymptomatic, mild, and severe, as well as outcomes such as hospitalization, ICU admission for critical illness, and mortality.

\subsection{Basic setup}

At the onset of the epidemic, the entire population is considered susceptible (S). Susceptible individuals become exposed and infected by the virus at a rate determined by the transmission rate parameter and the number of infectious and susceptible individuals in the population. Following a latent period, exposed individuals (E) transition to the infectious stage without symptoms, categorized as either completely asymptomatic (As) or presymptomatic (P). Upon completing the incubation period, presymptomatic individuals may develop either mild (Sm) or severe (Ss) symptoms. Asymptomatic individuals or those with mild symptoms eventually recover (R), while those with severe symptoms require hospitalization (H) after spending some time in the Ss compartment. Individuals requiring hospital care may either recover without complications or progress to critical illness (C) necessitating ICU admission. Those in critical condition may recover by transitioning back to post-ICU hospitalization (Hp) or succumb to the illness (D). All exposed individuals remain undetected until they transition to a symptomatic state. Upon entering asymptomatic, presymptomatic, mild, or severe states, a fraction of the individuals are detected after a certain period, leading to isolation and reduced infectiousness. A schematic representation of the model is provided in Figure 1. Model parameters, such as the incubation period, disease severity and symptomaticity, and lengths of hospital and ICU stays, are determined based on existing literature and observed data from Chicago. The transmission rate, probability of detection, and disease recovery rate are dynamically adjusted over time to better align with daily ICU and hospital census as well as reported deaths in Chicago. Further details regarding the model and its parameters, along with the source code, can be accessed from the GitHub repository \cite{covidmodel}.

\begin{figure}[!t]
    \centering
    \includegraphics[width = 0.5\textwidth]{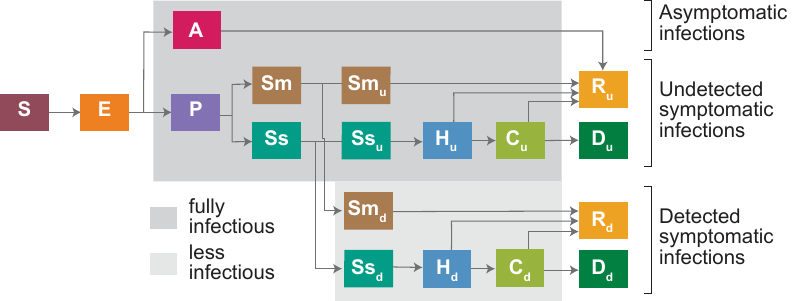}
    \caption{{\bf SEIR model to simulate SARS-CoV-2 transmission.}
 S=Susceptible, E=Exposed, A=Asymptomatic, P=Presymptomatic, Sm=Mild symptomatic, Ss=Severe symptomatic, H=Requires hospitalization, C=Critically ill (ICU), D=Death, R=Recovered. Subscripts u and d denote if the infection is undetected or detected. Asymptomatic and presymptomatic infections may also be detected, which would result in less infectiousness.}
    \label{fig:covid-age}
\end{figure}

\subsection{Checkpointing}
We have implemented checkpointing in the model such that the current state of the model (i.e., the number of persons in each state, the future state transition events, the current simulated time, etc.) can be serialized to a file and then restarted from that file, continuing from that serialized state. More importantly, when restarting from a checkpoint, the serialized state can be updated from user-input, parameterizing the model along a new trajectory. The following parameters can be specified to create a new trajectory: 1) the random seed; 2) the fraction of persons moving from E to P; 3) fraction of persons moving from P to Sm; 4) infectiousness of persons with symptomatic versus asymptomatic infections; 5) infectiousness of persons with detected versus undetected infections; 6) rate of persons moving from S to E. For illustrations in this paper, we treat the last parameter as the calibration parameter and set the other parameters to values obtained from prior study.

Checkpointing plays a pivotal role within the proposed inference framework, with detailed insights provided in subsequent sections. Here, we briefly explore its significance and advantages within sequential Monte Carlo algorithms. Within a sequential inference paradigm, where model parameters evolve over time, the simulation model must be executed iteratively to accommodate these parameter updates. For instance, when simulating an epidemic, the dynamic nature of the process demands continuous adjustment of model parameters with respect to time. By preserving the detailed state of the model at intermediate time points through checkpointing, we enable the seamless updating and application of parameters during forward simulation. This preserves the historical data and obviates the need to restart the simulation from the epidemic's onset.

% kaysmp(false), kmild(false), frac_as(false), frac_det(false), ini_ki(false), ki_ap(false)

\section{Statistical Model Formulation}
\label{sec:method}

% parameters should be made time varying in eq 1, 2, 3

The application involves a stochastic compartmental disease simulator (i.e., a computer model) that takes a $d$ dimensional input $\thetab = (\theta_1, \ldots, \theta_d)$ and produces vector output $\etabT{1}{T} = (\eta_1, \ldots, \eta_t)$, indexed by time $t$. The outputs are characteristics of the disease dynamics, such as the number of infectious individuals $\etabT{1}{T}^c$ and deaths $\etabT{1}{T}^d$, some of which are (partially) observable in the real world. In the case of stochastic simulation, each model run also requires the specification of a random seed $s$ to constitute a one-to-one mapping between input $(\thetab, s)$ to $\etabT{1}{T}$. We note that this level of input specification is one of the novel ingredients of our calibration framework. Trajectory outputs from an epidemic simulation, even at the same input values represent specific histories and dynamics of the disease, which are otherwise lost when represented via summary statistics. We assume that observed data include the reported number of cases -- $\ybT{1}{T}^c$ and deaths -- $\ybT{1}{T}^d$ during a disease outbreak. These data series are used to calibrate the stochastic disease simulation, i.e., to estimate $\thetab | (\ybT{1}{T}^c, \ybT{1}{T}^d)$. 

The basic observation model for each output type is written as:
\begin{equation}\label{eq:general}
    \ybT{1}{T} = \etabT{1}{T}(\thetab, s) + \epsb_T, \;\; \epsb_T  \sim MVN(0, \Sigma_T^2)
\end{equation}
i.e., we assume that the empirical counts are noisy observations of the simulated trajectories at an unknown ($\thetab, s)$. The likelihood $l(\ybT{1}{T} | \thetab, s)$ following the model is simply a multivariate Gaussian with mean $\etabT{1}{T}(\thetab, s)$ and a diagonal covariance matrix. The posterior $\pi(\thetab, s | \ybT{1}{T})$ is then proportional to $l(\ybT{1}{T} | \thetab, s) \times \pi(\thetab, s)$, where, $\pi(\thetab, s)$ denotes the joint prior on $\thetab$ and $s$. This posterior can then be studied by repeatedly sampling from it via common Monte-Carlo techniques such as MCMC, importance sampling, and accept-reject methods. One can produce summary statistics and credible intervals of $\thetab$ and future predictions using the samples from the posterior distribution. In this paper, we adopt a sequential Monte Carlo (SMC) approach based on importance sampling. 

\subsection{Bias model}
When calibrating a computer model, a bias term is often used to represent the misalignment between the true physical process and the computer simulation, which is often limited by our understanding of the true underlying system or the fidelity of the numerical solver. This is more commonly referred to as \emph{discrepancy} and is used as an additive or multiplicative factor to \ref{eq:general}. Similarly, in the case of an epidemic application, there are sources of discrepancy such as inaccurate reporting of cases and reporting lag, which can potentially affect the estimation of the model parameters and their uncertainties if ignored. Next we describe our proposed bias model with a focus on epidemic applications.

In the early days of an epidemic when data collection systems are often at an immature stage, the case counts can suffer from a lack of reporting. To capture this \textit{missing} count process, we use a binomial model to represent the observed counts as a fraction of the true unobserved counts. The basic underlying assumption is that for each occurrence of an event, it gets observed with probability $0< \rho_t < 1, t \in \{1, 2, \cdots \}$. Then, the total number of reported counts at time $t$, $\yb_t$ follows a binomial distribution with probability $\rho_t$. We write, $\yb_t \sim \text{Binomial}(\yb_t^{\text{true}}, \rho_t)$. In practice, both $\yb_t^{\text{true}}$ and $\rho_t$ are unknown and our goal is to infer $\yb_t^{\text{true}}$. The observation model in \ref{eq:general} for each time index $t$ can now be rewritten as:
\begin{eqnarray}\label{eq:general_w_bias}
    \yb_t = \etabo_t(\thetab, s, \rho_t) + \epsb_t, \;\; \varepsilon_t \sim N(0, \sigma_t^2), 0<\rho_t<1
\end{eqnarray}
where, $\etabo_t(\thetab, s, \rho_t) \sim \text{Binomial}(\etab_t(\thetab, s), \rho_t)$, i.e., the simulated observed counts $\etabo_t$ is a binomial random variable depending on the true simulated counts $\etab_t$ and the unknown probability $\rho_t$. The posterior is accordingly updated by modifying the likelihood that reflects the updated mean of the Gaussian distribution.
\begin{align}  
\label{eq:likelihood}
   &   l(\ybT{1}{T} | \thetab, s, \rho_t)  =  
     \frac{1}{\sqrt{2 \pi} |\Sigma_T|^{1/2}} \times   \\ \nonumber
     & \qquad \exp \Big\{ -\frac{1}{2}\left(\ybT{1}{T} - \etaboT{1}{T}(\thetab, s, \rho_t)^T\right) \\ \nonumber
     & \qquad\qquad\qquad\qquad \Sigma_T^{-1} \left(\ybT{1}{T} - \etaboT{1}{T}(\thetab, s, \rho_t)\right)\Big\}  
\end{align}
Together with the prior specification for $\thetab$, $s$, and $\rho_t$, the posterior can be fully obtained and sampled from. For the rest of the paper, we assume that $\rho_t$ does not change significantly within a relatively shorter time window, i.e., $\rho_t = \rho$.

\subsection{Sequential Importance Sampling (SIS)}
\label{sec:sis}
Importance sampling is a Monte Carlo technique used to estimate properties of a probability distribution, typically used when sampling directly from that distribution is difficult or inefficient. Instead of directly sampling from the target distribution, which might be computationally expensive or intractable, importance sampling generates samples from a different, easier-to-sample-from distribution called the ``proposal distribution.'' These samples are then weighted according to how likely they would have been generated from the target distribution. The key idea is to bias the sampling towards regions of the target distribution where the function being integrated or the property being estimated is large, thereby reducing the variance of the estimate. By adjusting for the differences in probabilities between the target and proposal distributions, importance sampling can provide more accurate estimates than direct sampling, especially in high-dimensional or complex scenarios.

Sequential importance sampling (SIS, \cite{doucet2001sequential,cappe2007overview}) is an extension of importance sampling that is particularly useful in scenarios where the target distribution changes over time, such as in an epidemic process. In SIS, instead of generating all samples at once from the target posterior distribution, samples are generated sequentially over time, with each sample's importance weight being updated as new information becomes available. This allows SIS to adaptively track changes in the target distribution. The typical workflow of SIS involves iteratively generating samples from the proposal distribution, updating their importance weights based on the likelihood of the observed data given the sample, and then resampling a subset of these samples according to their weights to prevent degeneracy (i.e., where only a few samples have significant weights). This process is repeated as new data arrive, allowing SIS to provide estimates of the target distribution that are conditioned on the observed data sequence. The steps to draw samples from $p(x_{0:t} \,|\, y_{1:t})$, which represents the probability of the state variables $x_{0:t}$ given the observed data sequence $y_{1:t}$ are as follows:
\begin{itemize}
    \item \textbf{Initialization}: Start with initial samples and weights drawn from the proposal distribution:
    \[ x_{0:t}^{(i)} \sim q(x_{0:t}) \quad \text{and} \quad w_t^{(i)} \]
    
    \item \textbf{Sequential Sampling and Weight Update}: At each time step \( t \), perform:
    \begin{enumerate}
        \item \textit{Sampling}:
        \[ x_t^{(i)} \sim q(x_t \,|\, x_{t-1}^{(i)}) \]
        
        \item \textit{Weight Update}:
        \[ w_t^{(i)} \propto w_{t-1}^{(i)} \frac{p(y_t \,|\, x_t^{(i)})}{q(x_t^{(i)} \,|\, x_{t-1}^{(i)})} \]
    \end{enumerate}
    
    \item \textbf{Normalization}:
    \[ w_t^{(i)} = \frac{w_t^{(i)}}{\sum_{j=1}^N w_t^{(j)}} \]
    
    \item \textbf{Resampling}: Resample $x_t^{(i)}$ with probabilities given by the weights $w_t^{(i)}$ to construct samples from the target distribution.
\end{itemize}

\subsection{Full Bayesian inference}
\label{sec:seq_inference}

Having reviewed all the necessary components, we provide a comprehensive outline of the sequential inference scheme applied to our epidemic system. To illustrate, let's consider a scenario resembling an actual epidemic. Imagine conducting a simulation over a total duration of $T$ weeks, during which it is compared against real-world data in discrete time windows, and model parameters are adjusted accordingly. We will describe this process for two such intervals, the first window and any window thereafter. Let $[1, t_1], \; t_1 > 0$ be the first time window and the $m$-th time window be denoted as $[t_{(m-1)}+1, t_m]$.

Following the model in \eqref{eq:general_w_bias}, the likelihood for the full $T$-variate observation $\ybT{1}{T} = (y_1, \ldots, y_T)$, can be decomposed into two independent components according to the window sizes. Furthermore, we assume that both case counts and the number of deaths are observable from the epidemic process and can be used to constrict the parameter tuple $(\thetab, s, \rho)$. The posterior of $(\thetab, s, \rho)$ conditioning on the full observation is as follows:

\begin{align}
\label{eq:full_post}
    & \pi\left(\thetab, s, \rho \big| \ybT{1}{T}^c, \ybT{1}{T}^d\right) \\ \nonumber
    &   \propto \;\;l\left(\ybT{1}{T}^c \big| \thetab, s, \rho\right) \times l\left(\ybT{1}{T}^d \big| \thetab, s, \rho\right) \times \pi(\thetab, s, \rho) \\ \nonumber
    % & \propto \exp \left\{ -\frac{1}{2}\left(\ybT{1}{T}^c - \etabT{1}{T}^c(\thetab, s, \rho)^T\right)\Sigma_t^{-1} \left(\ybT{1}{T} - \etabT{1}{T}^c(\thetab, s, \rho)\right)\right\}
    & \propto  \;\; \text{MVN}\left(\ybT{1}{T}^c \big| \etabT{1}{T}^c(\thetab, s, \rho), \Sigma_t^c\right) \times \\ \nonumber
    & \;\;\;\;\; \;\; \text{MVN}\left(\ybT{1}{T}^d \big| \etabT{1}{T}^d(\thetab, s, \rho), \Sigma_t^d\right) \times \\ \nonumber
    &  \;\;\;\;\;\;\; \pi(\thetab) \times \pi(s) \times \pi(\rho)
\end{align}
SIS is used to sample from this posterior and the posterior predictions for $\ybT{(T+1)}{T'}$ can be constructed by sampling from the Gaussian distribution with mean $\etab(\hat{\thetab}, \hat{s}, \hat{\rho})$ and covariance matrix $\hat{\Sigma}$.

\subsubsection{Inference in time window $[1:t_1]$}
The posterior for the first time window is trivially obtained from \eqref{eq:full_post}, by replacing $T$ by $t_1$. To sample from this posterior, we choose the prior distribution to be the same as the proposal distribution for importance sampling. In practice, this choice amounts to sampling from the prior distributions, running the simulations, and computing the importance weights by evaluating the likelihood function. 
\begin{algorithm}[!h]
\caption{Pseudo-code for single window IS}\label{alg:IS}
\begin{algorithmic}[1]
\REQUIRE  $N$ (total simulation budget)
\STATE Sample: $(\thetab_i, s_i, \rho_i) \sim \pi(\thetab, s, \rho), \;\;\; i=1:N$.
\STATE Run simulation: $\etabT{1}{t_1}^c(\thetab_i, s_i, \rho_i), \etabT{1}{t_1}^d(\thetab_i, s_i, \rho_i)$.
\STATE Compute weight: \\
\[w_{i}^{(t_1)} \propto \;\; l\left(\ybT{1}{t_1}^c \big| \thetab_i, s_i, \rho_i\right) \times l\left(\ybT{1}{t_1}^d \big| \thetab_i, s_i, \rho_i\right)\]
\STATE Resample: Sample  $(\thetab_1, s_1, \rho_1), \ldots, (\thetab_N, s_N, \rho_N)$ with probabilities proportional to $w_{i}^{(t_1)}$.
\end{algorithmic}
\end{algorithm}
At the end of the resampling step, the posterior epidemic trajectories are already available from the pool of simulation runs.

% \afnote{add a figure here}

\subsubsection{Inference in time window $[t_{(m-1)}+1, t_m]$}
To sample from $\pi\left(\thetab, s, \rho \; \big| \; \ybT{1}{t_m}^c, \ybT{1}{t_m}^d\right)$, i.e., the posterior conditioning on the full trajectory until time $t_m$, we make use of the posterior samples from $\pi\left(\thetab, s, \rho \; \big| \; \ybT{1}{t_{(m-1)}}^c, \ybT{1}{t_{(m-1)}}^d\right)$ as described in \ref{sec:sis}. For the sake of simplicity, we show the sequential update based on the conditional probability of the full posterior on the first two consecutive time windows, $[1, t_1]$ and $[t_1+1, t_2]$.
\begin{align}
    & \pi\left(\thetab, s, \rho \big| \ybT{1}{t_2}^c, \ybT{1}{t_2}^d\right) = \frac{l\left(\ybT{1}{t_2}^c, \ybT{1}{t_2}^d \big| \thetab, s, \rho\right) \pi\left(\thetab, s, \rho \right)}{\pi\left(\ybT{1}{t_2}^c, \ybT{1}{t_2}^d \right)} \\ \nonumber
    & \;\;\;\;\;\;  = \frac{l\left(\ybT{1}{t_1}^c, \ybT{1}{t_1}^d \big| \thetab, s, \rho\right) \pi\left(\thetab, s, \rho \right)}{\pi\left(\ybT{1}{t_1}^c, \ybT{1}{t_1}^d \right)} \;\; \bigtimes \\ \nonumber
    & \;\;\;\;\;\;\;\;  \frac{l\left(\ybT{(t_1+1)}{t_2}^c, \ybT{(t_1+1)}{t_2}^d \big| \ybT{1}{t_1}^c, \ybT{1}{t_1}^d, \thetab, s, \rho\right)}{\pi\left(\ybT{(t_1+1)}{t_2}^c, \ybT{(t_1+1)}{t_2}^d \big| \ybT{1}{t_1}^c, \ybT{1}{t_1}^d \right)}\\ \nonumber
    & \;\;\;\;\;\; \propto \;\; \pi\left(\thetab, s, \rho \big| \ybT{1}{t_1}^c, \ybT{1}{t_1}^d\right) \;\; \bigtimes \\ \nonumber
    & \;\;\;\;\;\;\;\;\;\; l\left(\ybT{(t_1+1)}{t_2}^c, \ybT{(t_1+1)}{t_2}^d \big| \ybT{1}{t_1}^c, \ybT{1}{t_1}^d, \thetab, s, \rho\right)
\end{align}
With the above formulation, samples from the intermediate posterior in the time window $1:t_1$ can now be used to update importance weights $w^{(t_2)}$ for the target posterior based on full $t_2$-variate observation. Replacing $t_1$ and $t_2$ by $t_{(m-1)}$ and $t_{m}$ for $m > 1$ will give the formula for the general time window $[t_{(m-1)}+1, t_m]$.

\subsubsection{Estimating time-varying epidemic parameters}
Apart from improving sampling efficiency, another rationale behind sequential calibration is to adapt to time-varying model parameters denoted as $\thetab_t$. Epidemic dynamics are subject to changing conditions, necessitating adjustments in model parameters over time. By calibrating the simulation model to real-world data within shorter time intervals sequentially, it becomes feasible to capture how model parameters evolve over time.

\section{Results}
\label{sec:results}

% why did we choose this (easy) ground truth scenario?

\subsection{Simulation setup}

To illustrate the sequential inference on the disease trajectory, an example from the Covid-age stochastic simulator is used as the ground truth empirical data. We use the transmission rate as the only model parameter that is to be estimated from the empirical data and this  parameter is allowed to vary over time, indicating an evolving epidemic. To make the problem of estimation of this time-varying parameter tractable, we assume that the value changes at discrete time points, called horizons, and the boundaries of the time windows for inference may or may not coincide with the horizons. To simulate the ground truth, we set the value of the transmission rate parameter at 0.3 for days 0 to 33,  0.27 for days 34 to 47,  0.25 from days 48 to 61, and 0.4 for days 62 onward. 
The selection of these values is qualitatively driven by the initial trends in COVID-19, where heightened awareness and social interventions lead to a decline in transmission rates over time, while also considering the potential emergence of new variants. The outputs from the simulator are the number of daily infections and deaths. The case counts from the simulator are treated as the actual number of cases, which is unobserved in reality. Hence, following the model in (\ref{eq:general_w_bias}), a bias is applied to the actual case counts to simulate observed case counts. The parameter of the binomial bias model is also set as a time-varying parameter with values of 0.6 for days 0 to 33,  0.7 for days 34 to 47, 0.85 for days 48 to 61, and 0.8 for days 62 and beyond. These values are chosen to mimic the improvement in data collection and reporting over time during a pandemic. 

\begin{figure}[!h]
    \centering
    \includegraphics[width = 0.5\textwidth]{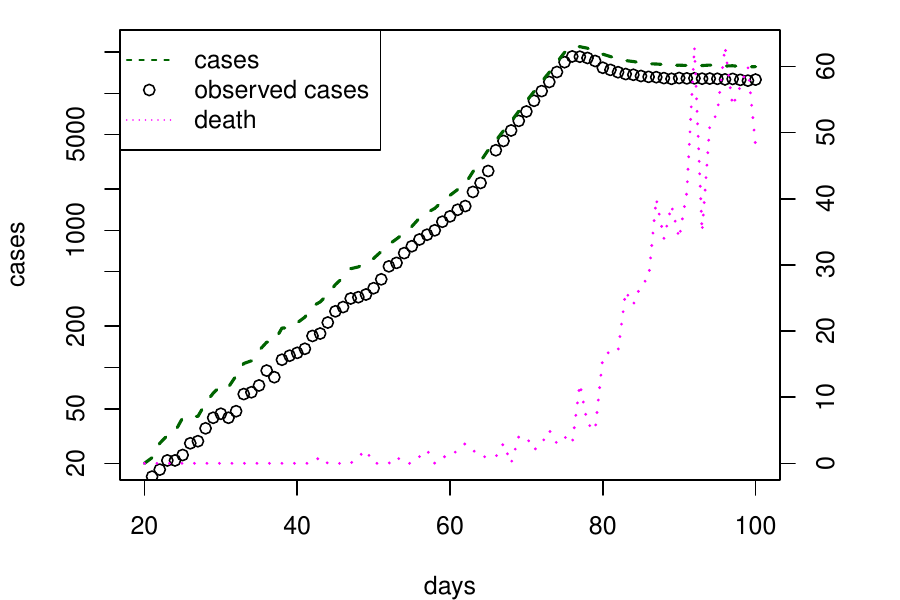}
    \caption{A simulated ground-truth using SEIR Covid model}
    \label{fig:gt}
\end{figure}

The sequential inference is performed in four time windows that reflect the changes in the epidemic behavior and the bias pattern. The steps to perform inference in each time window are the same, i.e., draw samples from the prior distribution of the 2-d parameter $(\theta, \rho)$ space, run simulations, compute the importance weight, and finally resample from the prior according to weights to construct the posterior. Except for the first window, the posterior samples from the previous window serve as the prior samples for the current window, as illustrated in \Cref{fig:four-stage-cases} and \Cref{fig:four-stage}. For the first time window, we use a uniform prior for the transmission rate parameter $\theta$ and a beta prior for the binomial bias model parameter $\rho$. At each unique parameter combination of $(\theta, \rho)$ the simulation is replicated 20 times.

\subsection{Calibrating to case counts}
\begin{figure*}[!h]
    \centering
    \includegraphics[width = 0.9\textwidth]{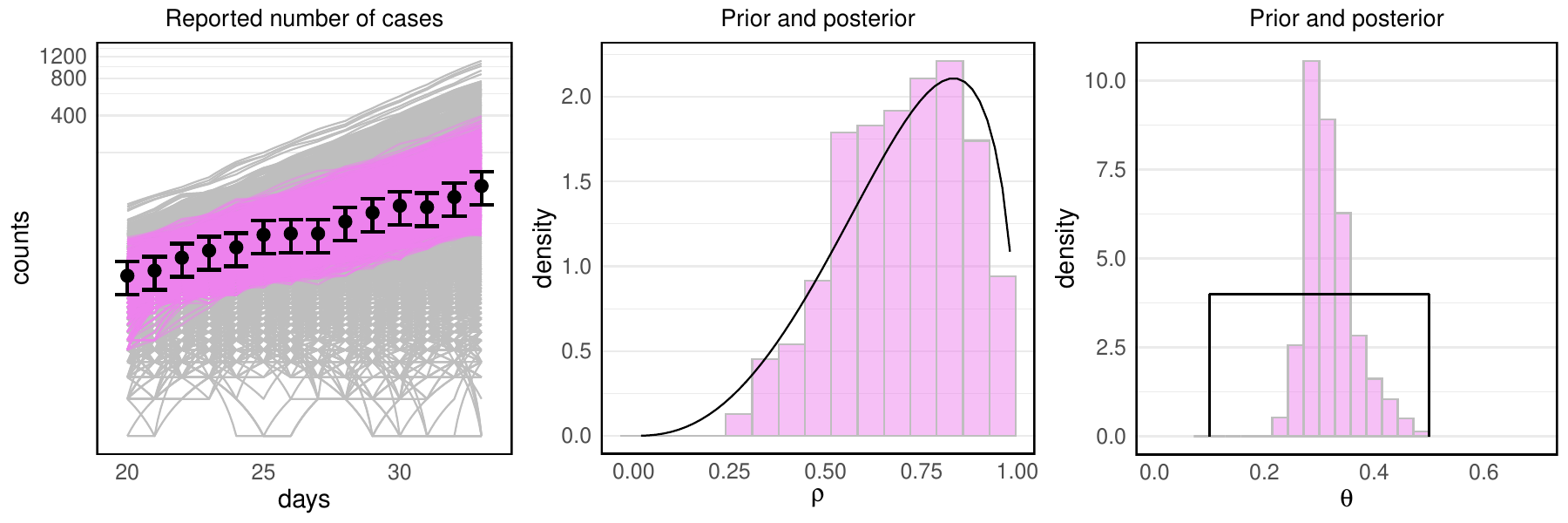}
    \caption{\textbf{Calibration using only reported case counts within the first time window} -- (left) The simulated ground truth is shown by black dots which are used to find the posterior trajectories (in purple) that are consistent with the ground truth from the prior trajectories (in grey), (center) the prior density (in black) and posterior of the $\rho$ parameter in the binomial bias model, (right) the prior and posterior distribution of simulation input $\theta$.}
    \label{fig:one-stage}
\end{figure*}

The first calibration experiment relies solely on the \textit{reported} case counts. Illustrated in \Cref{fig:one-stage}, this experiment demonstrates a single-step calibration employing importance sampling. Uniform priors over the interval $(0.1, 0.5)$ for $\theta$ and a $\beta(4, 1)$ prior for $\rho$ are considered, yielding a set of 25,000 prior samples. It's important to note that in the absence of prior information, an independent product prior is assumed for $(\theta, \rho)$. Since at each combination of $(\theta, \rho)$ the simulation is repeated 20 times, this generates a total of 500,000 unique trajectories. To control variability between replicates, the same set of random seeds is employed to generate the 20 realizations from the stochastic simulation. The left panel in \Cref{fig:one-stage} displays all 500,000 prior trajectories in grey. Using a Gaussian likelihood on square-root transformed counts with $\sigma_t = 1$, a sample of size 10,000 is then drawn with replacement from the 500,000 prior samples. This selection is based on probabilities proportional to the importance weights, thereby constructing the posterior distribution of the model parameters and output. The purple trajectories depicted in \Cref{fig:one-stage} represent the posterior model output, while the empirical histograms depict the posterior distributions on model parameters. It's worth noting that the posterior on $\rho$ exhibits less influence compared to that on $\theta$, partly due to the utilization of strong informative priors on $\rho$.

\begin{figure*}[!h]
    \centering
  \subfloat[\label{1a} Reported case counts (in black circles) are used to constrain the model parameters resulting in the posterior model output shown by grey lines. Purple and cyan shaded regions show 50\% and 90\% posterior credible intervals on the model output. The left plot shows the posterior on reported case counts and the right plot shows the posterior prediction on actual case counts that are unobserved. The blue dashed line represents the median predicted actual case counts. The vertical dashed grey lines represent the time boundaries for each calibration time window.]{%
       \includegraphics[width=0.6\linewidth]{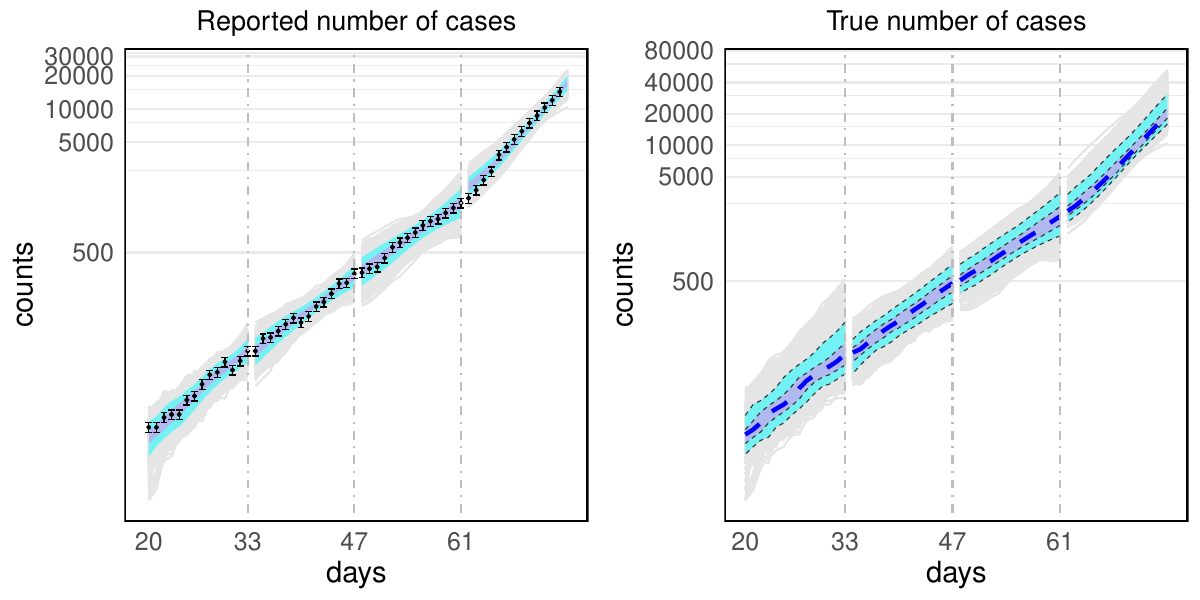}}
    \hfill
  \subfloat[\label{1b} 2d contour plot of the joint posterior of model parameters $(\theta, \rho)$ for four time windows respectively for which sequential calibration is performed. Black square denotes the true value of the parameters using which the ground truth is generated.]{%
        \includegraphics[width=0.99\linewidth]{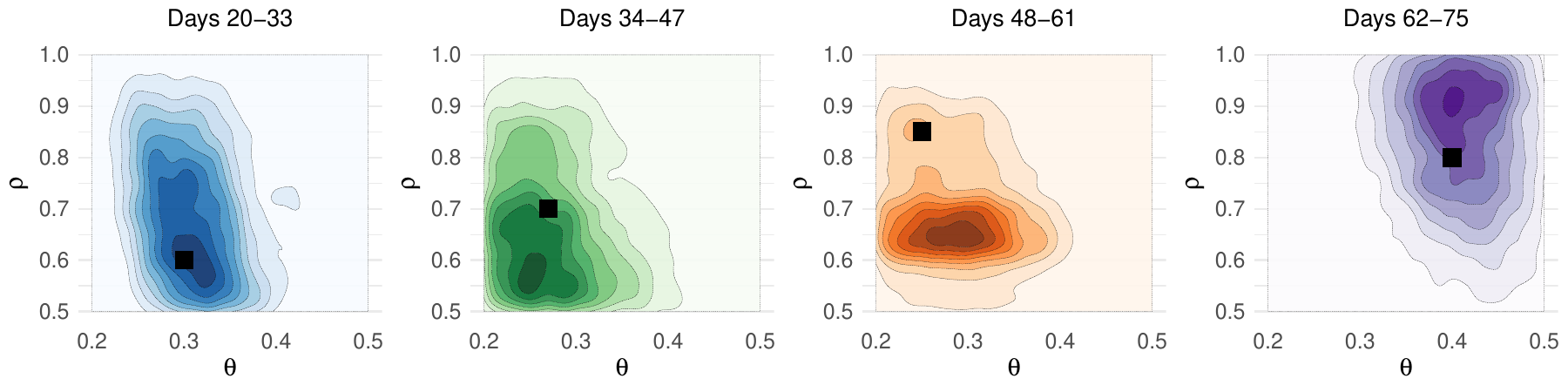}}
  \caption{\textbf{Sequential calibration using only reported case counts across four time windows}}
  \label{fig:four-stage-cases} 
\end{figure*}

Following the inference procedure described in \ref{sec:seq_inference}, \Cref{fig:four-stage-cases} illustrates the outcome of a comprehensive sequential calibration performed across four successive time intervals. Following the initial calibration stage (days 20-33), posterior models are preserved (i.e., checkpointed) to enable their forward simulation with updated model parameter values. Specifically, a uniform distribution centered around each posterior value of the model parameter is employed to generate prior samples for subsequent time intervals. Additionally, employing an informative prior, such as an asymmetric interval for the uniform distribution, may enhance the efficiency of importance sampling if such information is accessible, for example, if it is expected that the identification of cases is improving. We adopt a symmetric interval in the uniform prior for $\theta$ and an asymmetric interval for $\rho$, with a higher density toward the higher value of $\rho$, reflecting the reduced reporting error in later epidemic stages. Following the acquisition of calibrated parameters, we can deduce the actual number of unobserved cases in reality. The top-left plot in \Cref{fig:four-stage-cases} shows the simulated reported cases as black circles, alongside posterior trajectories in grey. The cyan and pink ribbons denote 50\% and 90\% credible intervals based on the posterior trajectories, respectively. By marginalizing out $\rho$ from the joint posterior of $(\theta, s, \rho)$, we can now formulate posterior predictive summaries and intervals on true case counts, showcased in the right plot. The bottom panel in \Cref{fig:four-stage-cases} illustrates the evolution of the posterior on model parameters as additional data is observed and utilized to constrain the trajectories over four time windows.

% \begin{figure*}[!t]
%     \centering
%     \includegraphics[width = 0.9\textwidth]{figs/combined_3_stage_post_all_cases.pdf}
%     \caption{\textbf{Sequential calibration using case counts} -- top panel: reported case counts (in black circles) are used to constrain the model parameters resulting in the posterior model output shown by grey lines. Pink and cyan shaded regions show 50\% and 90\% posterior credible intervals on the model output. The left plot shows the posterior on reported case counts and the right plot shows the posterior prediction on actual case counts that are unobserved. The black solid line represents the median predicted actual case counts. The vertical dashed grey lines represent the time boundaries for each calibration time window. Bottom panel: 2d contour plot of the joint posterior of model parameters $(\theta, \rho)$ for three time windows respectively for which sequential calibration is performed.}
%     \label{fig:three-stage-cases}
% \end{figure*}

\subsection{Calibrating to case counts and deaths}

\begin{figure*}[!ht]
    \centering
  \subfloat[\label{2a} Reported case counts (in black circles) and deaths (in black triangles) are used to constrain the model parameters resulting in the posterior model output shown by grey lines. Purple and cyan shaded regions show 50\% and 90\% posterior credible intervals on the model output. From left to right, the plots show the posterior on reported case counts, actual case counts that are unobserved, and death counts respectively. The blue dashed line in the middle plot represents the median predicted actual case counts. The vertical dashed grey lines represent the time boundaries for each calibration time window.]{%
       \includegraphics[width=0.9\linewidth]{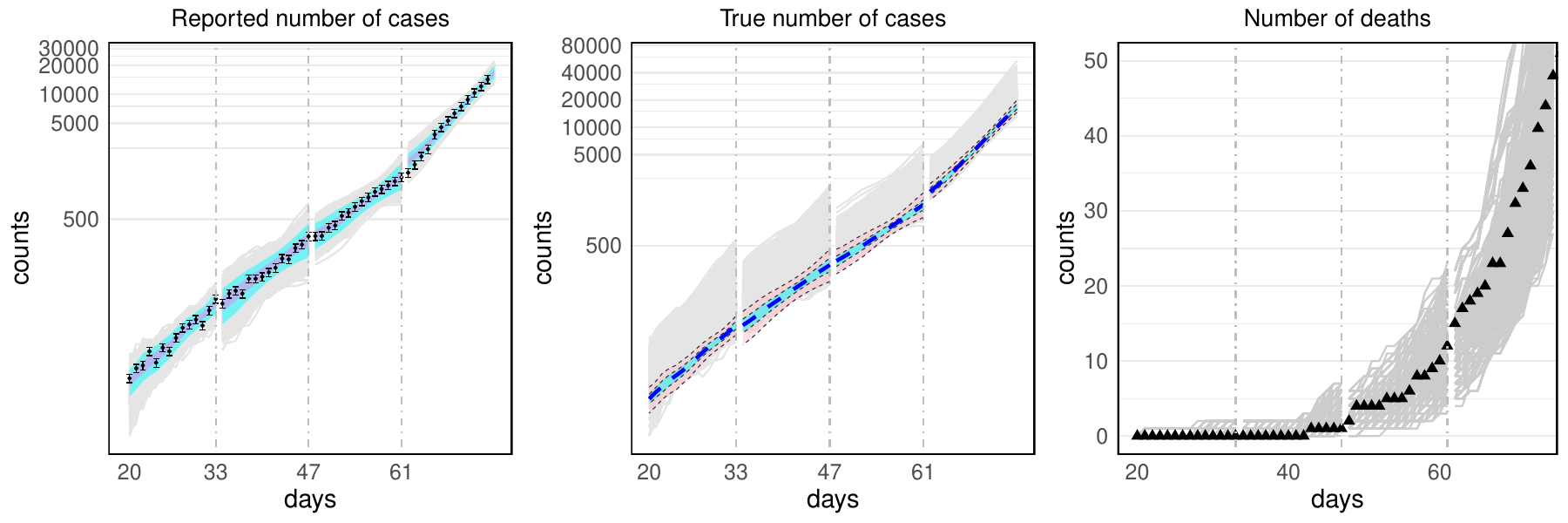}}
    \hfill
  \subfloat[\label{2b} 2d contour plot of the joint posterior of model parameters $(\theta, \rho)$ for four time windows respectively for which sequential calibration is performed. Black square denotes the true value of the parameters using which the ground truth is generated.]{%
        \includegraphics[width=0.99\linewidth]{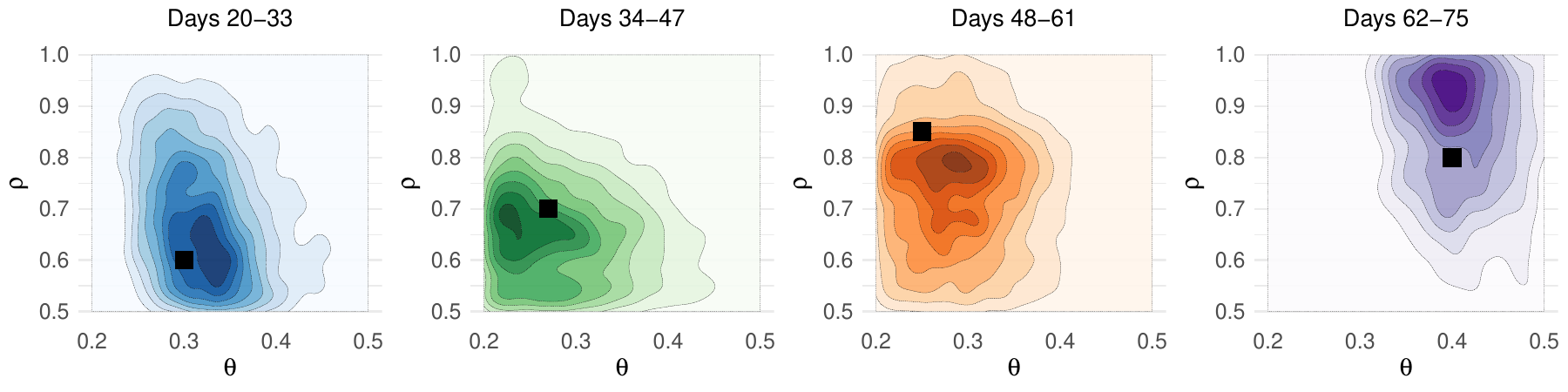}}
  \caption{\textbf{Sequential calibration using reported case counts and deaths across four time windows}}
  \label{fig:four-stage} 
\end{figure*}

In an effort to improve our time-varying estimation of $(\theta, \rho)$, we next integrate death counts as an additional model output. The model parameters are now constrained by both reported case counts and deaths. We do not assume any reporting bias on death counts, instead we use a Gaussian error model on the square-root counts similar to reported case counts. The updated posterior prediction, shown in \Cref{fig:four-stage}, encompasses reported cases, actual cases, and deaths. Notably, there is a reduction in uncertainty regarding reported case predictions. This trend is further reflected in the joint posterior contour plot, where high probability regions exhibit greater concentration around specific values. while our example operates under the assumption of no reporting bias in death counts, our SMC framework can easily be extended to incorporate bias in multiple data sources. It's worth emphasizing that this framework is highly adaptable, capable of incorporating various types of likelihoods, measurement bias models, and simulators, and can effectively handle multiple data sources and time windows.

% \begin{figure*}[!t]
%     \centering
%     \includegraphics[width = 0.9\textwidth]{figs/combined_3_stage_post_all.pdf}
%     \caption{\textbf{Sequential calibration using case and death counts} -- top panel: reported case and death counts (in black circles) are used to constrain the model parameters resulting in the posterior model output shown by grey lines. Pink and cyan shaded regions show 50\% and 90\% posterior credible intervals on the model output. The left plot shows the posterior on reported case counts and the middle plot shows the posterior prediction on actual case counts that are unobserved. The black solid line represents the median predicted actual case counts. The vertical dashed grey lines represent the time boundaries for each calibration time window. Bottom panel: 2d contour plot of the joint posterior of model parameters $(\theta, \rho)$ for three time windows respectively for which sequential calibration is performed.}
%     \label{fig:three-stage}
% \end{figure*}

\section{discussion}
\label{sec:discussion}
This model-based approach demonstrates the value of considering multiple types of information to constrain uncertainty when producing epidemic forecasts of difficult to observe quantities such as cases. This is accomplished via direct comparison of model outputs with surveillance data produced over the course of an epidemic.  For some outputs, there is a need to account for the form of known biases and errors; the simple Bernoulli-based error model that maps modeled cases to reported cases is an example of this -- we don't specify the actual probability a case is reported, but rather a simple, parameterized, probabilistic mechanism for reporting errors. As additional surveillance data are to be considered in the future, we'll need to develop models that link the simulated output (or perhaps model parameters) to the actual observation to be used. 

This SMC approach produces plausible epidemic trajectories/histories given the observed data.  While a compartmental model is used here, the approach applies equally well to other stochastic simulation models, such as ABMs. The computational demands of ABMs will likely require better efficiency; the use of surrogates for the individual trajectories (e.g. \cite{fadikar2023trajectory}) may be required to refine this current SMC implementation.  Unlike more aggregated models, these individual-based models provide a ``coordinate system'' that more readily maps to reality.  Hence, geographically and/or demographically targeted interventions, such as closing schools or vaccinating specific age-groups in certain areas, can be more easily captured in this type of modeling.  The trajectories produced from this SMC-based analysis can produce samples of plausible outcomes that allow direct, probabilistic assessment of different intervention strategies.  Still, these interventions will need to be encoded into these individual-based models in order to simulate their impact on future trajectories.
    
We note there is still some work to be done to ensure this approach remains operational as we consider more complicated epidemic models over many time periods. 
Our example contained only a few parameters for each time period, and even after considering multiple time periods, the resulting parameter space was still small compared to parameter spaces typical of large, geophysical systems \cite{anderson1999monte}. 
It's known that SMC has difficulties in settings with large parameter
spaces \cite{bengtsson2008curse}. Therefore, as this effort advances we'll also integrate reduced model and dimension reduction techniques \cite{BennerCOW2017ModRedApproxTA, wycoff2022sensitivity, wycoff2021sequential} to ensure that our approach remains feasible as we consider more complex modeling and additional time periods.

Another challenge is to ensure the posterior ensemble of model trajectories remains close to the actual epidemic. Our work uses stochastic simulations, which alleviates the potential issue of posterior weights concentrating on just a few draws from the importance distribution. However, if even the most highly weighted trajectories don't track reality, the SMC will produce unreliable predictions.  
Allowing the parameters to vary over time and allowing for error models to link model output to reality makes it easier for our model to produce trajectories that align with reality.  Also, producing larger numbers of trajectories in our importance sampler will combat this potential challenge.  Hence there will be a need for experimental platforms (e.g., \cite{collier_developing_2023,scalable_epi}) to diagnose such problems and develop potential fixes. A fuller consideration of these issues is relegated to future work.

\bibliographystyle{plain}
\bibliography{refs}

\end{document}